\documentstyle[fleqn,iopfts,12pt]{ioplppt}
\jl{1}
\eqnobysec
\begin{document}
\title{The $\bdelta$-deformation of the Fock space}
\author{Krzysztof Kowalski and Jakub Rembieli\'nski}
\address{\it Department of Theoretical Physics, University
of \L\'od\'z, ul. Pomorska 149/153,\\ 90-236 \L\'od\'z, Poland}
\begin{abstract}
A deformation of the Fock space based on the finite difference
replacement for the derivative is introduced.  The deformation
parameter is related to the dimension of the finite analogue
of the Fock space.
\end{abstract}
\section{Introduction}
In recent years there has been a growing interest to discretizations
of quantum mechanics based on the finite difference replacement for
the derivative.  This is motivated by the well-known speculations
that below the Planck scale the conventional notions of space and
time break down and the new discrete structures are likely to
emerge.  This has echoes in the arguments put forward in string
theory and quantum gravity.  We also mention the technical reasons
for the application of discrete models.  Let us only recall the
lattice gauge theories.  As a matter of fact the connection has been shown
in ref.\ 1 between ordinary quantum mechanics on a equidistant
lattice, where the the role of the derivative is played by the
forward or backward discrete derivative, and {\em q}-deformations
utilizing the Jackson derivative, nevetheless no explicit form of
the corresponding deformation of the Fock space has been provided in
ref.\ 1.  On the other hand, there are indications \cite{2} that
approaches based on the central difference operator are more
adequate for discretization of quantum mechanics than those using
asymmetric forward or backward discrete derivatives.

In this paper we introduce a deformation of the Fock space, such
that the creation and annihilation operators are elements of the
quotient field of the deformed Heisenberg algebra generated by the
usual position operator and the central difference operator.  The
deformation parameter $\delta$ describing the fixed coordinate
spacing is naturally related to the dimension of the
finite-dimensional space which can be regarded as an analogue
of the Fock space.  In the formal limit $\delta\to 0$ we arrive at
the infinite-dimensional space coinciding with the usual Fock space.
\newpage
\section{The $\bdelta$-deformation of the Heisenberg algebra}
As mentioned in the introduction there are indications that
discretizations of quantum mechanics should involve the central
difference operator such that
\begin{equation}
\Delta_\delta f(x)=\frac{f(x+\delta)-f(x-\delta)}{2\delta}.
\end{equation}
Furthermore, it seems to us that the most natural candidate for the
position operator in any discretized version of quantum mechanics is
the standard one of the form
\begin{equation}
\hat xf(x)=xf(x).
\end{equation}
In order to close the algebra satisfied by the operators
$\Delta_\delta $ and $\hat x$ we introduce the operator $I_\delta $
defined by
\begin{equation}
I_\delta f(x)=\frac{f(x+\delta )+f(x-\delta )}{2}.
\end{equation}
It follows that
\begin{equation}
[\Delta_\delta ,\hat x]=I_\delta,\qquad [I_\delta ,\hat
x]=\delta^2\Delta_\delta ,\qquad [I_\delta ,\Delta_\delta ]=0.
\end{equation}
Evidently,
\begin{equation}
\Delta_\delta=\hbox{$\scriptstyle{\rm i}\over\delta $}\sin\delta\hat p,\qquad I_\delta
=\cos\delta\hat p,
\end{equation}
where $\hat p=-{\rm i}\frac{d}{dx}$ is the usual momentum operator,
so the contraction of the algebra (2.4) referring to $\delta\to 0$, is
the usual Heisenberg algebra
\begin{equation}
[\hat x,\hat p]={\rm i}I.
\end{equation}
Using (2.1), (2.2) and (2.3) we find easily the following Casimir
operator for the algebra (2.4):
\begin{equation}
I_\delta^2-\delta^2\Delta_\delta^2=1.
\end{equation}

We now discuss the representations of the algebra (2.4).  We first observe
that (2.4) can be related to the following deformation of the $e(2)$
algebra (A.3) (see appendix):
\begin{equation}
[J,U_\delta ]=\delta U_\delta ,
\end{equation}
where $U_\delta $ is unitary, by means of the relations such that
\numparts
\begin{eqnarray}
\hat x &=& J,\\
\Delta_\delta &=& -\hbox{$\scriptstyle1\over2\delta$}(U_\delta
-U_\delta^\dagger),\\
I_\delta &=& \hbox{$\scriptstyle1\over2$}(U_\delta + U_\delta^\dagger).
\end{eqnarray}
\endnumparts
Consider the representation of (2.8) spanned by eigenvectors of the Hermitian
operator $J$.  Taking into account (2.8) and (A.5) we find
\begin{equation}
J|j\delta \rangle=j\delta  |j\delta \rangle.
\end{equation}
Hence, with the help of (2.8) we get
\begin{equation}
U_\delta  |j\delta \rangle= |(j+1)\delta \rangle,\qquad
U_\delta^\dagger |j\delta \rangle= |(j-1)\delta \rangle.
\end{equation}
Equations (2.9)--(2.11) taken together yield
\numparts
\begin{eqnarray}
\hat x |j\delta \rangle &=& j\delta  |j\delta \rangle,\\
\Delta_\delta  |j\delta \rangle &=& -\hbox{$\scriptstyle1\over2\delta
$}( |(j+1)\delta \rangle- |(j-1)\delta \rangle),\\
I_\delta  |j\delta \rangle &=& \hbox{$\scriptstyle1\over2$}(
|(j+1)\delta \rangle+ |(j-1)\delta \rangle).
\end{eqnarray}
\endnumparts
Let us now specialize to the case with integer $j$ (see appendix).
In view of the form of eq.\ (2.12a) it turns out that the operator
$\hat x$ really describes the position of a particle on equidistant
lattice with the fixed coordinate spacing $\delta $.  The
completeness condition satisfied by the vectors $ |j\delta\rangle$
can be written as
\begin{equation}
\sum_{j=-\infty}^{\infty}\delta  |j\delta \rangle\langle j\delta |=I.
\end{equation}
The relation (2.13) leads to the realization of the
abstract Hilbert space of states specified by the inner product
\begin{equation}
\langle f|g\rangle=\sum_{j=-\infty}^{\infty}\langle f|j\delta \rangle
\langle j\delta
|g\rangle\delta=\sum_{j=-\infty}^{\infty}f^*(j\delta)g(j\delta)\delta ,
\end{equation}
where $f(j\delta)=\langle j\delta|f\rangle$.  The action of operators
in the representation (2.14) is of the following form:
\numparts
\begin{eqnarray}
\hat xf(j\delta) &=& j\delta f(j\delta),\\
\Delta_\delta f(j\delta) &=& \hbox{$\scriptstyle1\over2\delta
$}[f((j+1)\delta)-f((j-1)\delta)],\\
I_\delta f(j\delta) &=& \hbox{$\scriptstyle1\over2$}[f((j+1)\delta)+
f((j-1)\delta)].
\end{eqnarray}
\endnumparts

We now study the representation generated by eigenvectors
$|\varphi\rangle_\delta $, $\varphi\in{\Bbb R}$,  of the unitary operator 
$U_\delta$ such that
\begin{equation}
U_\delta |\varphi\rangle_\delta =e^{-{\rm i}\delta\varphi}
|\varphi\rangle_\delta .
\end{equation}
It follows immediately from (2.9) and (2.16) that
\numparts
\begin{eqnarray}
\Delta_\delta  |\varphi\rangle_\delta &=& \hbox{$\scriptstyle{\rm
i}\over \delta $}\sin\delta \varphi |\varphi\rangle_\delta ,\\
I_\delta  |\varphi\rangle_\delta &=& \cos\delta \varphi
|\varphi\rangle_\delta .
\end{eqnarray}
\endnumparts
The completeness of the vectors $ |\varphi\rangle_\delta $ can be
expressed by
\begin{equation}
\frac{1}{2\pi}\int\limits_{-\frac{\pi}{\delta }}^{\frac{\pi}{\delta }}
 |\varphi\rangle_\delta {}_\delta\langle \varphi |=I.
\end{equation}
The resolution of the identity (2.18) gives rise to the functional
representation of vectors
\begin{equation}
\langle f|g\rangle =\frac{1}{2\pi}\int\limits_{-\frac{\pi}{\delta }}^{\frac{\pi}{\delta }}
f^*(\varphi)g(\varphi)d\varphi,
\end{equation}
where $f(\varphi)=\langle\varphi|f\rangle$, and we have omitted for
brevity the dependence of $f(\varphi)$ on $\delta $.  The operators
act in the representation (2.19) as follows:
\numparts
\begin{eqnarray}
\hat xf(\varphi) &=& {\rm i}\frac{d}{d\varphi}f(\varphi),\\
\Delta_\delta f(\varphi) &=& \hbox{$\scriptstyle {\rm i}\over\delta $}
\sin\delta\varphi f(\varphi),\\
I_\delta f(\varphi) &=& \cos\delta\varphi f(\varphi).
\end{eqnarray}
\endnumparts
Our purpose now is to analyze the contraction $\delta\to0$ of the
representations (2.14) and (2.19) introduced above.  Taking into
account (2.16), (2.13) and (2.11) we find that the passage from the
representation spanned by the vectors $ |j\delta\rangle$ and that
generated by the vectors $ |\varphi\rangle_\delta$ can be described
by the kernel
\begin{equation}
\langle j\delta |\varphi \rangle_\delta = e^{{\rm i}j\delta \varphi}.
\end{equation}
Equations (2.18) and (2.21) taken together yield
\begin{equation}
\langle j\delta|j'\delta\rangle =
\frac{1}{2\pi}\int\limits_{-\frac{\pi}{\delta }}^{\frac{\pi}{\delta }}
e^{{\rm i}(j-j')\delta\varphi}d\varphi =
\frac{\sin\pi(j-j')}{\pi(j-j')\delta}.
\end{equation}
Therefore
\begin{equation}
\langle j\delta|j'\delta\rangle = \hbox{$\scriptstyle
1\over\delta$}\delta_{jj'},
\end{equation}
whenever $\delta\ne 0$.  On the other hand, defining the continuum
limit as
\begin{equation}
j\to\infty,\qquad \delta\to0,\qquad j\delta = {\rm const} = x,
\end{equation}
and using the well known formula on the Dirac delta function
\begin{equation}
\delta(x) = \lim_{\alpha\to\infty}\frac{1}{\pi}\frac{\sin\alpha x}{x},
\end{equation}
we find that (2.22) takes the form
\begin{equation}
\lim\limits_{\scriptstyle j,\,j'\to\infty,\,\,\delta\to0
\atop\scriptstyle j\delta=x,\,j'\delta=x'}\langle j\delta|j'\delta\rangle
=\delta(x-x').
\end{equation}
Hence, we get
\begin{equation}
\lim\limits_{\scriptstyle j\to\infty,\,\,\delta\to0
\atop\scriptstyle j\delta=x} |j\delta\rangle =  |x\rangle,
\end{equation}
where $ |x\rangle$, $x\in{\Bbb R}$, are the usual normalized
eigenvectors of the position operator for a quantum mechanics on a
real line.  This observation is consistent with the fact that for
$\delta\to0$ the sum from (2.14) is simply the integral sum for the
scalar product in $L^2({\Bbb R},dx)$.  By (2.15) and (2.24) it is
also evident that in the limit $\delta\to0$ we arrive at the
Heisenberg algebra (2.6).  We have thus shown that the
contraction referring to $\delta\to0$ of the representation of the
algebra (2.4) given by (2.14) and (2.15) coincides with the standard
coordinate $L^2$ representation of the Heisenberg algebra (2.6).
Analogously, we have
\begin{equation}
{}_\delta\langle\varphi |\varphi'\rangle_\delta =
\sum_{j=-\infty}^{\infty}e^{-{\rm i}j\delta(\varphi -\varphi')}\delta.
\end{equation}
Therefore,
\begin{equation}
\lim_{\delta\to0}{}_\delta\langle\varphi
|\varphi'\rangle_\delta=2\pi\delta(\varphi -\varphi'),
\end{equation}
and we can identify
\begin{equation}
\lim_{\delta\to0} |\varphi\rangle_\delta = \sqrt{2\pi} |p\rangle,
\end{equation}
where $p=\varphi$, and $|p\rangle$, $p\in{\Bbb R}$, are the normalized
eigenvectors of the momentum operator.  Further, in view of (2.20)
the case $\delta\to0$ really corresponds to the Heisenberg algebra
(2.6).  So the representation specified
by (2.19) coincides in the limit $\delta\to0$ with the standard momentum
representation.  We conclude that the introduced deformation works
both on the level of the algebra and the representation.
\section{The $\bdelta$-deformation of the Heisenberg-Weyl algebra}
In this section we study the $\delta$-deformation of the
Heisenberg-Weyl algebra satisfied by the Bose creation and annihilation
operators.  Let us introduce the following family of operators:
\begin{equation}
A(s)=\hbox{$\scriptstyle1\over\sqrt{2}$}[\hat
x+(1-\delta^2s)\Delta_\delta I_\delta^{-1}],\qquad A^\dagger (s)=
\hbox{$\scriptstyle1\over\sqrt{2}$}[\hat
x-(1-\delta^2s)\Delta_\delta I_\delta^{-1}],
\end{equation}
where $s=0,1,\ldots.$  Clearly, these operators reduce to the
standard Bose creation and annihilation operators in the limit
$\delta\to0$.  We point out that then $A(s)$ and $A^\dagger(s)$ do
not depend on $s$.  Notice that in view of (2.9) $A^\dagger(s)$ is really 
the Hermitian conjugate of $A(s)$.  It should also be noted that in
the representation (2.20) the action of the operator $I_{\delta}^{-1}$ is
simply the multiplication by ${\rm sec}\,\delta\varphi$.  We now
seek the vectors $ |s\rangle$ and functions $\alpha(s)$ and
$\beta(s)$, satisfying
\begin{equation}
A(s) |s\rangle=\alpha(s) |s-1\rangle,\qquad A^\dagger(s)
|s\rangle=\beta(s) |s+1\rangle,\qquad s=0,1,\ldots .
\end{equation}
In other words, we are looking for the $\delta$-deformation of
vectors spanning the occupation number representation.  Using the
following form of the Casimir (2.7) which can be obtained with the
help of (3.1):
\begin{equation}
A(s+1)A^\dagger(s)-A^\dagger(s-1)A(s)=(1-\delta^2s)I,
\end{equation}
where $I$ is the unit operator, we get
\begin{equation}
\alpha(s+1)\beta(s)-\alpha(s)\beta(s-1)=1-\delta^2s.
\end{equation}
Hence, setting $\alpha(0)=0$ and solving the elementary recurrence
(3.4) we obtain
\begin{equation}
\alpha(s)\beta(s-1)=s-\hbox{$\scriptstyle\delta^2\over2$}s(s-1).
\end{equation}
The following solution of (3.5) consistent with the limit values
$\alpha(s)=\sqrt{s}$ and $\beta(s)=\sqrt{s+1}$, corresponding to
$\delta=0$, when $ |s\rangle$ span the usual occupation number
representation can be guessed easily:
\begin{equation}
\alpha(s)=\sqrt{s-\hbox{$\scriptstyle\delta^2\over2$}s(s-1)},\qquad
\beta(s)=\sqrt{s+1-\hbox{$\scriptstyle\delta^2\over2$}s(s+1)},
\end{equation}
so we have
\begin{equation}
A(s) |s\rangle=\sqrt{s-\hbox{$\scriptstyle\delta^2\over2$}s(s-1)}
|s-1\rangle,\qquad A^\dagger(s) |s\rangle=\sqrt{s+1-
\hbox{$\scriptstyle\delta^2\over2$}s(s+1)} |s+1\rangle.
\end{equation}
Now, by virtue of
\begin{equation}
\langle
s|A^\dagger(s)A(s)|s\rangle=[s-\hbox{$\scriptstyle\delta^2\over2$}
s(s-1)]\langle s-1|s-1\rangle\ge0,
\end{equation}
we see that the sequence of $s$ and thus $ |s\rangle$ should
truncate.  The only possibility left is to set
\begin{equation}
\delta^2=\frac{1}{s_{\rm max}}.
\end{equation}
Indeed, by (3.1) we then have
\begin{equation}
A(s_{\rm max})=A^\dagger(s_{\rm
max})=\hbox{$\scriptstyle1\over\sqrt{2}$}\hat x.
\end{equation}
Using this and (3.7), we find
\begin{equation}
 |s_{{\rm max}}+1\rangle= |s_{{\rm max}}-1\rangle,
\end{equation}
where $ |s_{\rm max}+1\rangle=A^\dagger(s_{\rm max}) |s_{\rm max}\rangle$.
We have thus shown that instead of $\delta$ we can use the parameter
$s_{{\rm max}}$ exceeding by one the dimension of the system of
vectors $\{|s\rangle\}_{0\le s\le s_{{\rm max}}}$.  Such systems for
$s_{\rm max}=1$, $s_{\rm max}=2$ and so on, can be interpreted as a
finite-dimensional analogues of the usual infinite-dimensional Fock space.  
The latter evidently refers to the case with $s_{\rm max}=\infty$, when
$\delta=0$.

We now discuss the algebra satisfied by the operators (3.1), that is
the $\delta$-deformation of the Heisenberg-Weyl algebra.  Taking
into account (3.7) we get
\numparts
\begin{eqnarray}
A(s')&=&\left[1-\frac{\delta^2(s-s')}{2(\delta^2s-1)}\right]A(s)+
\frac{\delta^2(s-s')}{2(\delta^2s-1)}A^\dagger(s),\\
A^\dagger(s')&=&\frac{\delta^2(s-s')}{2(\delta^2s-1)}A(s)+
\left[1-\frac{\delta^2(s-s')}{2(\delta^2s-1)}\right]A^\dagger(s),
\qquad s<s_{\rm max}.
\end{eqnarray}
\endnumparts
Making use of (2.4), (3.1), (3.12) and the following form of the Casimir
(2.7), which can be easily derived with the help of (3.1):
\begin{equation}
\delta^2[A(s)-A^\dagger(s)]^2=2(1-\delta^2s)(1-I_\delta^{-2}),
\end{equation}
we arrive at the commutation relations such that
\numparts
\begin{eqnarray}
&&\fl [A(s),A^\dagger(s')]=[1-\hbox{$\scriptstyle\delta
^2\over2$}(s+s')]I_\delta^{-2},\\
&&\fl [A(s),A(s')]=[A^\dagger(s'),A^\dagger(s)]=\hbox{$\scriptstyle\delta
^2\over2$}(s'-s)I_\delta^{-2},\qquad s,s'\le s_{\rm max},\\
&&\fl [A(s),I_\delta^{-2k}]=[A^\dagger(s),I_\delta^{-2k}]=[A(s_{\rm max}),
I_\delta^{-2k}]=k\frac{\delta^2}{1-\delta^2s}B_k(s),\quad s<s_{\rm max},\\
&&\fl [A(s),B_k(s')]=[A^\dagger(s),B_k(s')]\nonumber\\
&&\fl\quad{}=2k(1-\delta^2s')I_\delta^{-2k}-
(2k+1)(1-\delta^2s')I_\delta^{-2(k+1)},\\
&&\fl [B_k(s),I_\delta^{-2l}]=[I_\delta^{-2k},I_\delta^{-2l}]=
[B_k(s),B_l(s')]=0,\,\, s,s'\le s_{\rm max},\,\, k,l=1,2,\ldots,
\end{eqnarray}
\endnumparts
where $B_k(s)=[A(s)-A^\dagger(s)]I_\delta^{-2k}$.
We remark that due to the commutator (3.14d) the algebra (3.14) is infinite
dimensional.  It should also be noted that in view of the following
relation:
\begin{equation}
A(s)=(1-\hbox{$\scriptstyle\delta^2s\over2$})A(0)+
\hbox{$\scriptstyle\delta^2s\over2$}A^\dagger(0),\qquad 0\le s\le
s_{\rm max},
\end{equation}
which is an immediate consequence of (3.1), $A(s)$, $A^\dagger(s)$ and
$B_k(s)$ can be regarded as a discrete curve in the algebra
generated by $A(0)$, $A^\dagger(0)$, $I_\delta^{-2k}$ and $B_k(0)$
of the form
\numparts
\begin{eqnarray}
&&[A(0),A^\dagger(0)]=I_\delta^{-2},\\
&&[A(0),I_\delta^{-2k}]=[A^\dagger(0),I_\delta^{-2k}]=k\delta^2B_k(0),\\
&&[A(0),B_k(0)]=[A^\dagger(0),B_k(0)]=2kI_\delta^{-2k}-
(2k+1)I_\delta^{-2(k+1)},\\
&&[B_k(0),I_\delta^{-2l}]=[I_\delta^{-2k},I_\delta^{-2l}]=[B_k(0),B_l(0)]=0,
\qquad k,l=1,2,\ldots .
\end{eqnarray}
\endnumparts
Of course, both (3.14) and (3.16) reduce to the Heisenberg-Weyl algebra
in the limit $\delta\to0$, that is $s_{\rm max}\to\infty$.
\section{The $\bdelta$-deformation of the Fock space}
We now discuss the $\delta$-deformation of the Fock space expressed
by (3.7) in a more detail.  We first observe that the generation of
the states $|s\rangle$, with $s\ge1$, from the ``vacuum vector''
$|0\rangle$ can be described with the help of the second equation of
(3.7) by
\begin{equation}
|s\rangle=\left(\prod\limits_{s'=0}^{s-1}
\frac{1}{\sqrt{s'+1-\hbox{$\scriptstyle\delta^2\over2$}s'(s'+1)}}
\right)A^\dagger(s-1)\cdots A^\dagger(1)A^\dagger(0) |0\rangle,\quad
0<s\le s_{\rm max}.
\end{equation}
The vectors $|s\rangle$ are not orthonormal.  In fact, using (3.12a)
with $s'=s+1$, and (3.7) we find
\begin{eqnarray}
&&\fl\fl\,\,\delta^2\sqrt{s+1-\hbox{$\scriptstyle\delta^2\over2$}s(s+1)}
\sqrt{s-\hbox{$\scriptstyle\delta^2\over2$}s(s-1)}\langle s-1|s+1\rangle
=-2(\delta^2s-1)[s+1-\hbox{$\scriptstyle\delta^2\over2$}s(s+1)]\langle
s|s\rangle\nonumber\\
&&\fl\fl\quad{}+[\delta^2+2(\delta^2s-1)][s+1-\hbox{$\scriptstyle\delta^2\over2$}
s(s+1)]\langle s+1|s+1\rangle.
\end{eqnarray}
Further, calculating the expectation value of the Casimir (3.13) in
the state $|s\rangle$ with the use of (3.14a) for $s=s'$, and taking
into account (4.2), we obtain
\begin{eqnarray}
&&\fl\delta^2\sqrt{s+1-\hbox{$\scriptstyle\delta^2\over2$}s(s+1)}
\sqrt{s-\hbox{$\scriptstyle\delta^2\over2$}s(s-1)}\langle s+1|s-1\rangle
=2(\delta^2s-1)
[s-\hbox{$\scriptstyle\delta^2\over2$}s(s-1)]\langle
s|s\rangle\nonumber\\
&&\fl\quad{} + [\delta^2-2(\delta^2s-1)][s-\hbox{$\scriptstyle\delta^2\over2$}
s(s-1)]\langle s-1|s-1\rangle.
\end{eqnarray}
Equating right-hand sides of (4.2) and (4.3) we finally arrive at the
following recursive formula on the squared norm of $|s\rangle$:
\begin{eqnarray}
&&2(\delta^2s-1)(2s+1-\delta^2s^2)\langle s|s\rangle+[\delta^2-
2(\delta^2s-1)][s-\hbox{$\scriptstyle\delta^2\over2$}s(s-1)]
\langle s-1|s-1\rangle\nonumber\\
&&\quad{}-[\delta^2+2(\delta^2s-1)][s+1-\hbox{$\scriptstyle\delta^2\over2$}
s(s+1)]\langle s+1|s+1\rangle=0,\qquad s\le s_{\rm max}.
\end{eqnarray}
A straightforward calculation shows that the recurrence (4.4) can be
written in a more convenient form such that
\numparts
\begin{eqnarray}
&&\fl\langle 1|1\rangle=\frac{2}{2-\delta^2}\langle 0|0\rangle,\\
&&\fl\langle s|s\rangle=\frac{\delta^2}{[2(\delta^2s-1)-\delta ^2][s-
\hbox{$\scriptstyle\delta^2\over2$}s(s-1)]}\sum_{s'=0}^{s-2}
(\delta ^2s'-1)\langle s'|s'\rangle\nonumber\\
&&\fl\quad{}+
\left(1+\frac{\delta^2[\delta^2(s-1)-1]}{[2(\delta^2s-1)-
\delta^2][s-\hbox{$\scriptstyle\delta^2\over2$}s(s-1)]}\right)
\langle s-1|s-1\rangle,\qquad 2\le s\le s_{\rm max}.
\end{eqnarray}
\endnumparts
Finally, eqs.\ (3.12a) and (3.7) taken together yield
\begin{eqnarray}
&&\fl\fl\,\sqrt{s-\hbox{$\scriptstyle\delta^2\over2$}s(s-1)}\langle s|s'\rangle
=\left[1-\frac{\delta^2(s'-s+1)}{2(\delta^2s'-1)}\right]
\sqrt{s'-\hbox{$\scriptstyle\delta^2\over2$}s'(s'-1)}\langle s-1|s'-1\rangle
\nonumber\\
&&\fl\fl\!{}+\frac{\delta^2(s'-s+1)}{2(\delta^2s'-1)}
\sqrt{s'+1-\hbox{$\scriptstyle\delta^2\over2$}s'(s'+1)}\langle s-1
|s'+1\rangle,\!\!\!\quad 0<s\le s_{\rm max},\!\!\!\!\!\quad 0\le s'
<s_{\rm max}.
\end{eqnarray}
The equations (4.5) and (4.6) form the closed system which enables to
calculate the inner product $\langle s|s'\rangle$ for arbitrary
$s,s'\le s_{\rm max}$.  In particular, utilizing the relation
\begin{equation}
\langle s|s+1\rangle=0,\qquad s\le s_{\rm max},
\end{equation}
implied by (4.6) and using recursively (4.6) we find that
\begin{equation}
\langle s|s'\rangle=0, \qquad s,s'\le s_{\rm max},
\end{equation}
where $s$ is even and $s'$ is odd.

We finally discuss the concrete realization of the introduced
$\delta$-deformation of the abstract Fock space in the
representation (2.19).  On using (2.20) and (3.7) we arrive at the
following system:
\numparts
\begin{eqnarray}
\left[\frac{d}{d\varphi}+(1-\delta^2s)\frac{1}{\delta}{\rm tg}\delta
\varphi\right]f_s(\varphi)&=&-{\rm i}\sqrt{2}\sqrt{s-\hbox{$\scriptstyle\delta^2\over2$}
s(s-1)}\,f_{s-1}(\varphi),\\
\left[\frac{d}{d\varphi}-(1-\delta^2s)\frac{1}{\delta}{\rm tg}\delta
\varphi\right]f_s(\varphi)&=&-{\rm i}\sqrt{2}\sqrt{s+1-\hbox{$\scriptstyle\delta^2\over2$}
s(s+1)}\,f_{s+1}(p),
\end{eqnarray}
\endnumparts
where $f_s(\varphi)=\langle \varphi|s\rangle$.  We remark that the
system (4.9) is the special case of the more general one
\numparts
\begin{eqnarray}
\left[\frac{d}{d\varphi}+k(s,\varphi)\right]f_s(\varphi)&=&-{\rm i}\mu(s)
f_{s-1}(\varphi),\\
\left[\frac{d}{d\varphi}-k(s,\varphi)\right]f_s(\varphi)&=&-{\rm i}\nu(s)
f_{s+1}(\varphi).
\end{eqnarray}
\endnumparts
It can be easily checked that (4.10) is equivalent to
\numparts
\begin{eqnarray}
\left[\frac{d}{dx}+k(s,x)\right]y_s(x)&=&\mu(s)
y_{s-1}(x),\\
\left[-\frac{d}{dx}+k(s,x)\right]y_s(x)&=&\nu(s)
y_{s+1}(x).
\end{eqnarray}
\endnumparts
The system (4.11) was studied by Jannussis {\em et al} \cite{3} in the
context of the generalization of the Infeld-Hull method of factorization
in the case of the harmonic oscillator.  Analyzing the compatibility of the
two second order differential equations implied by (4.11) they showed
that besides the periodic solution there exists the following one:
\begin{equation}
k(s,x)=a{\rm ctg}(ax+\theta)\,s-\frac{b}{a}{\rm ctg}(ax+\theta)+
\frac{c}{\sin(ax+\theta)},
\end{equation}
provided
\begin{equation}
\mu(s)\nu(s-1)=-a^2s(s-1)+2bs+\lambda,
\end{equation}
where $a$, $b$, $c$, $\theta$ and $\lambda$ are are arbitrary
constants.  A look at (4.12), (4.13), (4.9) and (3.5) is enough to
conclude that the actual treatment refers to the case with
$a=\delta$, $b=1$, $c=0$, $\theta=\pi/2$ and $\lambda=0$.  We point
out that within the formalism introduced herein the second order
equations implied by (4.9) are simply the realization of the
abstract equations
\numparts
\begin{eqnarray}
A(s+1)A^\dagger(s)|s\rangle &=& \beta(s)\alpha(s+1)|s\rangle,\\
A^\dagger(s-1)A(s)|s\rangle &=& \alpha(s)\beta(s-1)|s\rangle.
\end{eqnarray}
\endnumparts
in the representation (2.19).  The compatibility of the eqs.\ (4.14)
is ensured by the Casimir (3.3).  In this sense the actual approach
can be interpreted as an abstract form of the Infeld-Hull factorization
method.

We now return to (4.9).  Using (4.9a) and the limit
\begin{equation}
\lim\limits_{\delta\to0}\,(\cos\delta \varphi)^{\frac{1}{\delta^2}}=
e^{-\frac{\varphi^2}{2}},
\end{equation}
we find
\begin{equation}
f_0(\varphi)=\pi^{-\frac{1}{4}}(\cos\delta \varphi)^{\frac{1}{\delta^2}}.
\end{equation}
Furthermore, utilizing (4.9b) and
\begin{equation}
\frac{d}{d\varphi}(\cos\delta \varphi)^{\frac{1}{\delta^2}}=-\frac{{\rm
tg}\delta \varphi}{\delta }(\cos\delta
\varphi)^{\frac{1}{\delta^2}},\qquad \frac{d}{d\varphi}\left(\frac{{\rm
tg}\delta \varphi}{\delta }\right)=1+\delta^2\left(\frac{{\rm
tg}\delta \varphi}{\delta }\right)^2,
\end{equation}
we get
\begin{equation}
f_s(\varphi)=\frac{\pi^{-\frac{1}{4}}(-{\rm i})^s}{(\sqrt{2})^s}
\left(\prod\limits_{s'=0}^{s-1}\frac{1}{\sqrt{s'+1-
\hbox{$\scriptstyle\delta^2\over2$}s'(s'+1)}}\right)
H^{(\delta)}_s\left(\frac{{\rm tg}\delta \varphi}{\delta}\right)
(\cos\delta \varphi)^{\frac{1}{\delta^2}},
\end{equation}
where $\quad 1\le s\le s_{\rm max}$, and $H^{(\delta)}_s(x)$ are
the polynomials satisfying the recurrence
\begin{eqnarray}
H^{(\delta)}_{s+1}(x)&=&(2-\delta^2s)xH^{(\delta)}_s(x)-(1+\delta^2x^2)
H^{(\delta ){}'}_s(x),\nonumber\\
H^{(\delta )}_0(x)&=&1,
\end{eqnarray}
where the prime designates the differentiation with respect to $x$.  Of
course, $H^{(\delta)}_s(x)$ are simply the $\delta$-deformation
of the usual Hermite polynomials refering to the limit $\delta\to0$,
i.e.\ $s_{\rm max}\to\infty$.  The first few $\delta$-deformed
Hermite polynomials are of the form
\begin{eqnarray}
\fl H^{(\delta)}_0(x)&=&1,\nonumber\\
\fl H^{(\delta)}_1(x)&=&2x,\nonumber\\
\fl H^{(\delta)}_2(x)&=&4(1-\delta^2)x^2-2,\nonumber\\
\fl H^{(\delta)}_3(x)&=&8(1-\delta^2)(1-2\delta^2)x^3-12(1-\delta^2)x,\nonumber\\
\fl H^{(\delta)}_4(x)&=&16(1-\delta^2)(1-2\delta^2)(1-3\delta^2)x^4
-48(1-\delta ^2)(1-2\delta ^2)x^2+12(1-\delta^2).
\end{eqnarray}
As with the standard Hermite polynomials the general formula on the
$\delta$-deformed ones can be derived such that
\begin{eqnarray}
\fl H^{(\delta)}_0(x)&=&1,\nonumber\\
\fl H^{(\delta)}_s(x)&=&\sum_{j=0}^{\left[\hbox{$\scriptstyle
s\over2$}\right]}
(-1)^j\frac{s!}{j!(s-2j)!}2^{s-2j}\left[\prod\limits_{s'=0}^{s-j-1}
(1-\delta^2s')\right]x^{s-2j},\qquad 1\le s\le s_{\rm max},
\end{eqnarray}
where $[y]$ is the biggest integer in $y$.

We finally write down
the following formula on the matrix elements $\langle s|s'\rangle$
implied by (2.18) and (4.18):
\begin{equation}
\langle s|s'\rangle =
\frac{1}{2\pi}\int\limits_{-\frac{\pi}{\delta}}^{\frac{\pi}{\delta}}
f_s^*(\varphi)f_{s'}(\varphi)d\varphi,
\end{equation}
where $f_s(\varphi)$ is given by (4.16) and (4.18).  The calculation
of the integral from (4.22) for arbitrary $s,\,s'$ seems to be more
complicated than the solution of the recurrences (4.5) and (4.6).
It should be noted however that (4.22) enables to calculate the
squared norm of the ``vacuum vector'' $ |0\rangle$ parametrizing
solutions of (4.5) and (4.6).  Namely, we find
\begin{equation}
\langle 0|0\rangle =
\frac{1}{2\pi^\frac{3}{2}}\int\limits_{-\frac{\pi}{\delta}}^{\frac{\pi}
{\delta}}
(\cos\delta\varphi)^\frac{2}{\delta^2}d\varphi = \sqrt{\frac{s_{\rm
max}}{\pi}}\,\frac{(2s_{\rm max}-1)!!}{(2s_{\rm max})!!}=
\frac{\sqrt{s_{\rm
max}}}{\pi}\,\frac{\Gamma(s_{\rm max}+\frac{1}{2})}{\Gamma(s_{\rm
max}+1)},
\end{equation}
where $\delta^2s_{\rm max}=1$ and $\Gamma(x)$ is the gamma
function.
\section{Conclusion}
We have introduced in this work the deformation of the Fock space
based on the utilization of the central difference operator instead
of the usual derivative.  It should be mentioned that there exist
alternative approaches for discretization of quantum mechanics
relying on finite difference representations of the usual Heisenberg
\cite{4} or Heisenberg-Weyl algebra \cite{5}.  Nevertheless, the general problem
with them is the interpretation of the nonequivalence of the
obtained representations of the canonical commutation relations and
the standard Schr\"odinger one.  Some problems with the spectrum of
operators within such approaches have been also reported \cite{4}.
We also recall the discretization of the harmonic oscillator
introduced in \cite{6} relying on the replacement of the Hermite
polynomials with the Kravchuk polynomials in a discrete variable
as well as the finite-dimensional counterpart of the Fock space
spanned by the eigenvectors of the phase operator discussed in
\cite{7}.  In analogy with the actual treatment in both approaches
taken up in \cite{6} and \cite{7} the standard infinite-dimensional
Fock space refers to the formal limit $N\to\infty$, where $N$ is 
dimension of the finite-dimensional discrete version of the Fock space.
Moreover, in the case with the discretization described in \cite{6} one can
recognize a counterpart of the parameter $\delta$ specified by (3.9)
such that $\delta\simeq N^{-\frac{1}{2}}$.  Nevertheless, besides of
those similarities we have also serious differences.  For example,
in opposition to the operators (3.1) the generalizations of the Bose
operators introduced in \cite{6} do not depend on the index
labelling the basis of the finite-dimensional analogue of the
Fock space.  On the other hand, the alternatives to the number
states discussed in \cite{7} form the orthonormal set.  This is not
the case for the states $ |s\rangle$ described herein.
Last but not least we point out that besides of quantum mechanics the
results of this paper would be of importance in the theory of
differential equations.  We only recall the abstract form of the
Infeld-Hull method of factorization described by the equations (4.14)
and (3.3).
\appendix
\section*{}
\renewcommand{\theequation}{A.\arabic{equation}}
Here we briefly discuss the basic properties of the $e(2)$ algebra.
Consider the $e(2)$ algebra
\begin{equation}
[J,X]={\rm i}Y,\qquad [J,Y]=-{\rm i}X,\qquad [X,Y]=0.
\end{equation}
The Casimir operator for (A.1) is of the following form:
\begin{equation}
X^2+Y^2=r^2.
\end{equation}
Making use of (A.1) and (A.2) we arrive at the following form of the
algebra (A.1):
\begin{equation}
[J,U]=U,
\end{equation}
where
\begin{equation}
U=\hbox{$\scriptstyle 1\over r$}(X+{\rm i}Y)
\end{equation}
is unitary.  Consider the eigenvalue equation
\begin{equation}
J |j\rangle = j |j\rangle.
\end{equation}
From equations (A.3) and (A.5) it follows that the operators $U$ and
$U^\dagger$ act on the vectors $ |j\rangle$ as the rising and
lowering operator, respectively, that is
\begin{equation}
U |j\rangle =  |j+1\rangle,\qquad U^\dagger |j\rangle =  |j-1\rangle.
\end{equation}
Taking into account (A.6) we find that the whole basis {$
|j\rangle$} of the Hilbert space of states can be generated from the
unique ``vacuum vector'' $ |j_0\rangle$, where $j_0\in[0,1]$.  The
non-equivalent irreducible representations of the commutation
relations (A.3) are labelled by different $j_0$.  We remark that the
algebra (A.3) is the most natural for the study of a quantum
particle on a circle \cite{8}.  In such a case $J$ represents the angular
momentum and the unitary operator $U$ describes the position of a
particle on a unit circle.  We now demand the time-reversal
invariance of the algebra (A.3).  Having in mind the interpretation
of $J$ as the angular momentum this leads to
\begin{eqnarray}
TJT^{-1} &=& -J,\\
TUT^{-1} &=& U^{-1},
\end{eqnarray}
where $T$ is the anti-unitary operator of time inversion.  Using
(A.5)--(A.8) we obtain
\begin{equation}
T |j\rangle =  |-j\rangle.
\end{equation}
As an immediate consequence of (A.9), we find that $T$ is well
defined on the Hilbert space of states generated by the vectors $
|j\rangle$ if and only if the spectrum of $J$ is symmetric with
respect to zero.  Hence, in view of (A.6) the only possibility left
is $j_0=0$ or $j_0=\frac{1}{2}$.  Obviously, $j_0=0$
($j_0=\frac{1}{2}$) implies integer (half-integer) eigenvalues $j$.

However, in this work we interpret $J$ as the
position operator for a quantum particle on a lattice.  Accordingly,
the operator $T$ from (A.7) should be replaced with a unitary parity
operator $P$ and the invariance of (A.3) under parity transformation
demanded.  In that case the relations (A.7) and (A.8) (with $T$
replaced by $P$) and their consequences ($j$ integer or half-integer)
remain unchanged.

\end{document}